\documentclass[twocolumn,showpacs,preprintnumbers,amsmath,amssymb]{revtex4}

\usepackage{graphicx}
\begin{document}

\newlength{\plotwidth}
\setlength{\plotwidth}{8.5cm}

\title{Nanomachining of multilayer graphene using an atomic force microscope}

\author{P.~Barthold}
\author{T.~L\"udtke}
\author{R.~J.~Haug}
\affiliation{Institut f\"ur Festk\"orperphysik, Leibniz
Universit\"at Hannover, Appelstr. 2, 30167 Hannover, Germany}
\date{\today}
\begin{abstract}
An atomic force microscope is used to structure a film of
multilayer graphene. The resistance of the sample was measured
\textit{in-situ} during nanomachining a narrow trench. We found a
reversible behavior in the electrical resistance which we
attribute to the movement of dislocations. After several attempts
also permanent changes are observed. Two theoretical approaches
are presented to approximate the measured resistance.
\end{abstract}
%73.63.b Electronic transport in nanoscale materials and structures
%73.22.-f Electronic structure of nanoscale materials: clusters, nanoparticles, nanotubes, and nanocrystals
%73.23.-b Electronic transport in mesoscopic systems
%81.07.-b Nanoscale materials and structures: fabrication and characterization
%PACS, the Physics and Astronomy Classification Scheme.

\maketitle

Atomic force microscopes (AFMs) are well known tools for imaging
and for structuring. Besides other lithographic methods
nanomachining with the AFM is a simple, but highly efficient way
to design devices on the sub-micron level. By applying a high
contact force between sample and AFM tip a permanent deformation
of the sample's surface is obtained. Using this method different
materials have been structured e.g.
semiconductors~\cite{Magno_apl_70,schumacher_apl_75,
Regul_APL_81} and metals~\cite{Irmer_apl_73}.\\
Up to now the common technique to structure graphene is by
etching.~\cite{Oezyilmaz_apl_91,stampfer_tunable_nanostructured,russo_condmat_ring}
Graphene has drawn a great deal of attention since the discovery
of free standing single layer graphite (so-called graphene) and
its unique electronic properties.~\cite{Novesolov_science_666,
kim_nature_438,Novoselov_science_315,geim_nature_mat_6} The
motivation for the work presented here was to structure graphene
via nanomachining with an AFM tip. We structured a
thin film of graphite by nanomachining a trench through the half
width of the sample. Hence the conducting area of the sample is
reduced and thereby a constriction is formed. Thereby we observed an
interesting reversible behavior in the resistance and in the end a
permanent change in the resistance.\\
The graphite sample used in this study is extracted from natural
graphite~\cite{graphit} by exfoliation \cite{Novoselov_PNAS_102}
on a silicon substrate with a 300~nm SiO$_\mathrm{2}$ layer. The
thereby formed flake has a lateral dimension of a few micrometers
and a thickness of about 10~nm ($\sim$30 atomic layers, assuming a
lattice constant of 0.34~nm). The Ti/Au (9~nm/46~nm) electrodes
are fabricated using standard electron beam lithography. After
bonding the sample it is electrically contacted inside the AFM
allowing \emph{in-situ} measurements at room temperature.
\begin{figure}
\includegraphics[scale=1]{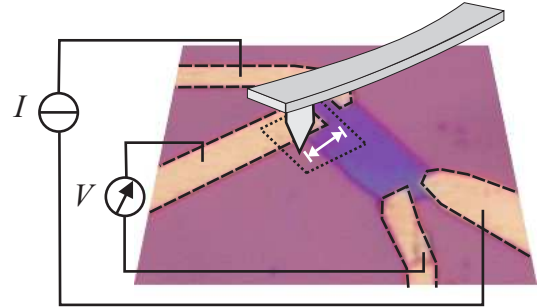}
\caption{\label{fig:setup} Schematic drawing of the setup. The
optical picture shows the graphite flake with four electrodes. A
direct current is driven through the sample via two contacts while
the voltage is measured using the other two electrodes. The AFM
tip moves from left to right and to the left again while a high
contact force is applied. The dashed square marks the region which
is shown in Fig.~\ref{fig:trench}(a) and (b).}
\end{figure}
Figure~\ref{fig:setup} shows the general setup. A direct current
of $I=500$~nA is driven via two contacts through the sample while
the voltage $V$ is measured using the two remaining electrodes.
For the measurements presented here we used an AFM tip that is
coated with polycrystalline diamond on the tip-side. During the
measurements we applied a force of approximately 0.5~$\mu$N. Using
such a high contact force the tip is moved with a velocity of
about 0.5~$\mu$m/s half the way across the graphite flake as
sketched by the white arrows in Fig.~\ref{fig:setup}. The tip
starts its movement left of the flake, moves about 2.2~$\mu$m
through the flake and returns back to its starting position. Thus
the tip scratches the sample in both directions. After five of
those movements a distinct trench
is formed in the graphite film.\\
Figures~\ref{fig:trench}(a) and (b) show two AFM pictures of the
sample before and after nanomachining. A trench of about
2.2~$\mu$m is clearly visible in the graphite flake in
Fig.~\ref{fig:trench}(b).
\begin{figure}
\includegraphics[scale=1]{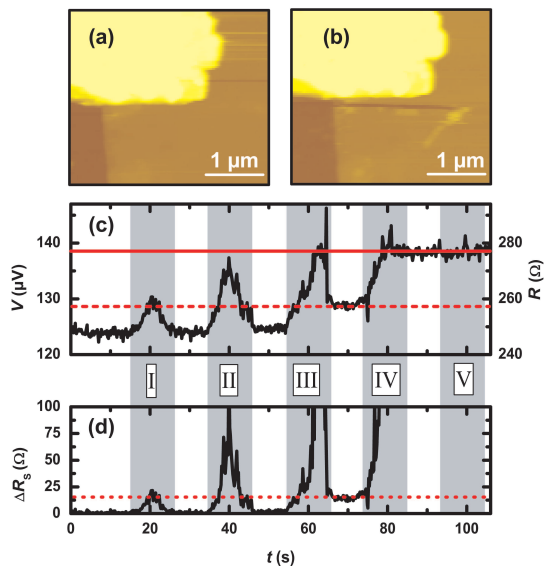}
\caption{\label{fig:trench}Upper part: AFM images of the graphite
flake with a height of about 10~nm. (a) Magnification of the
interesting part as marked in Fig~\ref{fig:setup}. (b) After
nanomachining five times with the AFM tip. A clear trench in the
graphite is visible. Its dimensions are
$w_\mathrm{S}\approx2.2~\mu$m and $l_\mathrm{S}\approx100$~nm. (c)
Time evolution of the resistance of the sample while the AFM tip
applies a force to the graphite. The grey regions indicate when
the tip actually moves on top of the graphite, the roman numerals
count the number of movements. The dashed and solid vertical lines
are guides to the eye to stress the similar resistances of the
sample during different times of structuring. (d) The resistance
change of the structured part $\Delta R_S$ is shown. $\Delta R_S$
is extracted from Fig.~\ref{fig:trench}(c) by subtracting
$R(t=0~$s) leading to $\Delta R_S = 0~\Omega$ and putting $\Delta
R_S (t=500~$s) to infinity as the graphite is cut through in this
part of the sample.}
\end{figure}
Figure~\ref{fig:trench}(c) demonstrates the time evolution of the
overall resistance $R$ while scratching the graphite film with the
AFM tip.  To demonstrate the resistance change of the structured
part $\Delta R_S$ is shown in Fig.~\ref{fig:trench}(d). The time
period when the tip moves on top of the graphite is marked grey in
Fig.~\ref{fig:trench}(c) and (d). At $t=0$~s the resistance of the
sample is about $R\approx248~\Omega$ ($\Delta R_S=0~\Omega$). At
$t\approx15$~s when the tip is moved over the graphite for the
first time (I) with the high contact force the resistance starts
to increase. The resistance reaches its first maximum of
$R\approx258~\Omega$ ($\Delta R_S\approx15~\Omega$) at
$t\approx21$~s which coincides with the reversal point of the AFM
tip movement. The resistance drops again to its original value by
moving the tip back to the original starting position. When the
tip applies a force to the graphite for the second time the
resistance starts to rise again~(II). This time the value rises up
to about 275~$\Omega$. Afterwards the resistance drops to a value
of 248~$\Omega$ ($\Delta R_S=0~\Omega$). As the AFM tip moves over
the flake for the third time (III) the resistance increases to a
value of about 277~$\Omega$. Now the resistance decreases to
$R\approx258$~$\Omega$, which is 10~$\Omega$ higher than the
overall resistance in the beginning and corresponds to a $\Delta
R_S\approx15~\Omega$. The value after the third tip movement is
the same as the maximum obtained during AFM run I, as indicated by
the dashed line in Fig.~\ref{fig:trench}(c). As the tip moves for
the fourth time (IV) over the graphite, the resistance rises again
to a value of about 277~$\Omega$. The same value is already
reached during run II and III. But this time the resistance does
not drop again instead it stays at a value of
$R\approx277~\Omega$. This resistance is kept even when the tip
moves for a fifth time (V) on top of the graphite and stays at
this value afterwards. Thus the resistance of the graphite film
was permanently changed by 29~$\Omega$ using an AFM tip to structure it.\\
\begin{figure}
\includegraphics[scale=1]{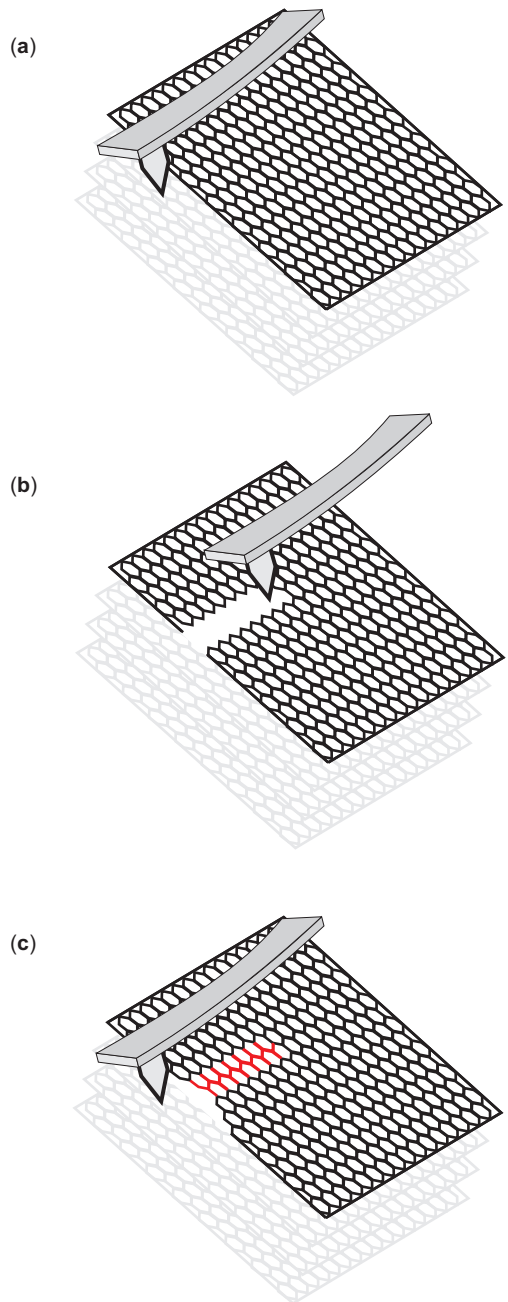}
\caption{\label{fig:trench_scheme}A simplified schematic sketch of
the AFM tip as the graphene layers are nanomachined. (a) The
sample is still unperturbed. (b) The tip has been moved over the
sample and bonds have been destroyed. (c) as the tip moves back
the induced dislocations start to move to the edge of the sample.}
\end{figure}
To explain this behavior we consider the following model: While
the AFM tip is moved over the sample dislocations are induced
along the trajectory of the movement of the tip as schematically
depicted in Fig.~\ref{fig:trench_scheme}(b). These dislocations
modify the electronic properties of the sample. Thus the
resistance of the sample rises during scratching. These
dislocations then move to the edge of the sample where we assume
that their influence on the electronic properties of the flake is
only small illustrated in Fig.~\ref{fig:trench_scheme}(c). Grenall
reported dislocation movement in smeared flakes of natural
graphite.~\cite{Grenall_nature_182} As observed by Williamson
dislocations in graphite run parallel to the layer
plane.~\cite{Williamson_proc_royal_257} Mainly they move to the
edge of the flake or to cleavage steps. Hence bonds just destroyed
by the AFM tip along the trajectory of the movement could close
again and the transport properties get back to the original state,
thus the resistance drops again to its original value.\\
As our sample consists of many layers, it seems reasonable to
believe that during the first time the sample is scratched (I)
dislocations are induced only in the few upper layers and during
the second time (II) dislocations are induced in more layers. This
would explain the higher resistance during run~II compared to I.
As the resistance during run~II is close to the value reached at
the end, dislocations seem to be formed in
most of the layers when scratching for the second time.\\
The defects induced during the second time of scratching could
move again to the edge of the sample. Therefore the resistance
drops (between II and III) to its original value. During the third
time of scratching (III) a lasting deformation occurs for the
first time. In a few layers the bonds destroyed by the AFM tip are
not closed again and thereby influence the electronic properties
of the sample permanently. During the fourth run (IV) all layers
are cut through on a 2.2~$\mu$m long path along the sample. Thus
the resistance keeps its value even when it is scratched for the
fifth time. All bonds are destroyed along the trajectory of the
movement of the AFM tip.\\
\begin{figure}
\includegraphics[scale=1]{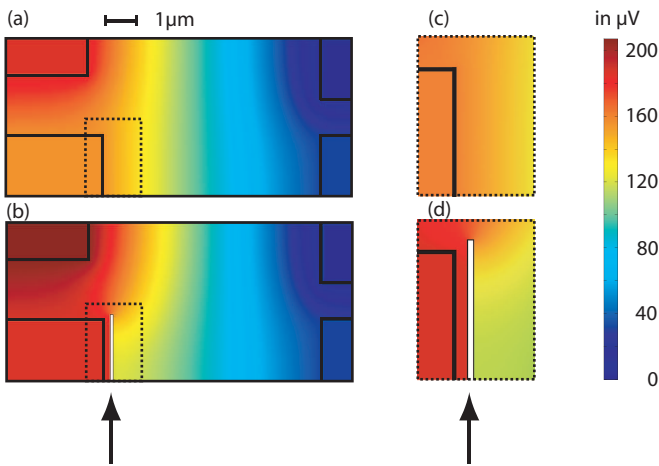}
\caption{\label{fig:epot}Top view of the numerical calculated
electrical potential of the sample. The arrow marks the
nanomachined trench. The gold electrodes are depicted by the black
rectangles on the edges of the sample. The current is driven
through the upper contacts while the lower ones are used to
measure the voltage drop. (a) The electrical potential drop of the
unperturbed sample. (b) Sample after five runs with the AFM and a
trench formed with the dimensions $w_\mathrm{S}\approx2.2~\mu$m
and $l_\mathrm{S}\approx100$~nm. A dramatic change in the
potential drop between the two lower electrodes is clearly
visible, (c) shows a magnification of the later structured part
and (d) demonstrates a blow up of the surrounding of trench.}
\end{figure}
What follows now are two theoretical approaches to get an
understanding of the nanomachining process in terms of  the
measured resistances. In a first step to model the resistances we
start with Ohm's law:
\begin{equation}
\label{eq:ohm} \mathbf{J}=\sigma\mathbf{E},
\end{equation}
where $\mathbf{J}$ is the current density, $\sigma = \rho^{-1}$
the conductivity tensor, $\mathbf{E}=-\nabla V$ is the electric
field with the potential $V$. Applying the conservation of
currents to Eq.~\ref{eq:ohm} leads to:
\begin{equation}
\label{eq:conti} \nabla \mathbf{J}=\nabla\sigma\mathbf{E}=
-\nabla\cdot(\sigma\nabla V)=0
\end{equation}
The current is driven through the upper left electrode in
Fig.~\ref{fig:epot}. The boundary conditions were selected to be
electrically insulating (the normal component of the current
density is zero, $\mathbf{n}\cdot\mathbf{J}=0$).The second order
partial differential Eq.~\ref{eq:conti} is numerically solved. The
calculations were performed using finite elements within in a mesh
of around 40,000 elements.\cite{comsol} This is a three
dimensional, diffusive model. Knowing the geometry of the sample
we find a sheet resistivity of $\rho \approx
2.03\cdot10^{-6}\Omega$m. The measured values are
$R\approx248~\Omega$, $l\approx7.3~\mu$m, $w\approx5.2~\mu$m, and
$h\approx10$~nm, where $l$ is the length in current direction, $w$
the width orthogonal to the current direction, and $h$ the height
of the sample. By applying these values to the textbook formula
$\rho=R \cdot w\cdot h/{l}$, the resulting resistivity is $\rho
\approx1.77\cdot10^{-6}~\Omega$m, being comparable to the one
found by our numerical calculations and the specific resistance
$\rho\approx1.2\cdot 10^{-6}~\Omega$m reported by Powell et al.
for natural graphite.~\cite{Powell_AIP_142} The reason for the
difference between these two latter results might be that the
resistivity of graphite depends strongly on the doping of the
sample and thereby varies from sample to sample and Powell et al.
report the specific resistances of samples that are in the
dimensions of a few millimeters and centimeters. The difference
compared to the numerical calculations is caused by the fact that
the textbook formula describes an ideal macroscopic system. In the
numerical model for our mesoscopic device the asymmetry in the
electrodes of our sample is taken into account. Therefore we will
use $\rho \approx 2.03\cdot10^{-6}\Omega$m for our
further calculations.\\
Figure~\ref{fig:epot} shows the results of the numerical
simulation. The electrical potential of the intact flake
Fig.~\ref{fig:epot}~(a) and the potential of the sample in the end
with the formed trench Fig.~\ref{fig:epot}~(b) are compared. If a
trench with $w_S = 2.2~\mu$m and $l_S = 100$~nm is simulated
within this numerical model a drastic change in the electrical
potential is clearly visible. Figure~\ref{fig:epot}~(c) and (d)
illustrate the dramatic influence of the relatively small trench
on the electrical potential of the sample. This radical change in
the potential leads to a resistance change between the lower
electrodes in Fig.~\ref{fig:epot} of 63~$\Omega$ (measured
resistance change 29~$\Omega$). As we are dealing here with a
mesoscopic device, reasons for the differences between the
measured and the calculated resistance change might be that
neither side-effects nor quantum effects are taken into account by
the simulation. In addition the nanomachnined trench is relatively
small compared to the size of the sample. The numerical model
simulates the whole sample and as it provides good results for the
global measured effects, we consider another theoretical approach
that describes the resistance change
from a more local point of view.\\
For this we compare our results to findings of Garc\'ia et
al.~\cite{Esquinazi} who used an exact evaluation of Maxwell's
solutions for a spreading, ohmic resistance of a constriction
separating two semi-infinite media. In this two dimensional model
the resistance value contributed by the formed trench in the end
can be described using Eq. 4 from~\cite{Esquinazi}, which could be
written in the following form for our problem:
\begin{equation}
\label{eq:esqu} R_{2d}=\frac{2a\rho}{h\pi}\cdot
\ln(\frac{w}{w-w_S}),
\end{equation}
where $R_{2d}$ is the spreading, ohmic resistance of the
constriction, $a$ is a constant that takes care of the influence
of the sample shape and the topology of the electrodes position,
and $w$ is the width of the sample, whereas $w_S$ is the width of
the structured part, hence $w-w_S$ is the width of the
constriction. As we are at room temperature ballistic parts are
negligible. Also the length of the constriction is much smaller
than the width, and therefore it also does not contribute. In our
device the most of the voltage drop measured by the electrodes is
given by the $7.3~\mu$m long path, therefore the increase in the
resistance due to the constriction can be estimated by using
Eq.~\ref{eq:esqu}. With $\rho=2.03\cdot 10^{-6}\Omega m$, $h =
10~$nm, $w=5.2~\mu$m, $w-w_s=3~\mu$m and $a=1/2$ it leads to
$R_{2d}\approx 35.5~\Omega$. With $R\approx 248~\Omega$ of the
unperturbed sample this results in $R = 283.5~\Omega$ in the end,
which compares nicely with our measured resistance $R \approx
277~\Omega$. As this model only describes the resistance change
due to the locally formed constriction, it is not dependent on the
geometry and homogeneity of the rest of the sample. Therefore it
is quite reasonable that this estimation fits better than the
numerical evaluation which depends on the whole sample. In a
perfectly shaped and homogeneous device both results should
converge to one another. Let us point out that both here used
methods to estimate the quantitative findings are rough
assumptions, based on the one hand side on a three dimensional,
numerical model and on the other hand side on a strictly two
dimensional, analytical approach.  An adequate model to describe
our device in more detail would be
needed.\\
In conclusion, we have shown \textit{in-situ} measurements of the
resistance of mesoscopic graphite being nanomachined with a
diamond coated AFM tip, for one exemplary sample, other measured
results can be found in Ref.~\cite{barthold_nanomachining}. During
processing the device we find a reversible change in the
electrical resistance. We attribute this effect to induced
dislocations that lead to an increased resistance. At room
temperature these dislocations can easily move to the edges of the
graphite flake leading to reversible resistance changes. After
processing the sample with the AFM tip a couple of times the
resistance changes permanently, i.e. bonds inside the graphite are
broken permanently. Two different theoretical models are
demonstrated to estimate the measured resistance changes. Further
investigations of the reversible resistance change should be
performed varying other parameters as for example the temperature
and the velocity of the AFM tip movement to see how the reversing
of the resistance depends on those parameters and to learn more
about the influence of the dislocation movement on the electronic
properties of graphene layers. Forming smaller constrictions the
observation of ballistic contribution in the transport should be
possible.~\cite{Esquinazi} The here presented technique to
nanomachine mesoscopic graphite with an AFM contributes to the
promising prospective approach to create a device based on single
layer graphene which has not yet been
successful.~\cite{Zeitler_nanolithography}

\bibliographystyle{prsty}

\begin{thebibliography}{10}
\bibitem{comsol}
The predifened enviroment for the dc-conductive media static case
of the commercial finite elements software COMSOL has been used to
solve this partial differential equation.
\end{thebibliography}

\end{document}